\documentclass[12pt]{article}
\usepackage{a4,latexsym,pslatex,graphicx,times,color,amsmath}
\usepackage{subfig} 

\usepackage[numbers]{natbib}

\title{Semi-analytic velocity profile for a Bingham fluid in a curved channel}
\author{T.G. Roberts and S.J. Cox \\ 
Department of Mathematics, Aberystwyth University, SY23 3BZ, UK.}

\begin{document}

\maketitle

\begin{abstract}
We derive an expression for the velocity profile of a pressure-driven yield-stress Bingham fluid flowing around a 2D concentric annulus. The formula requires the numerical solution of a nonlinear equation for the positions of the yield surfaces. The results allow the prediction of the effects of channel curvature on the pressure gradient required to initiate flow for given yield stress, and for the width of the plug region and the flux through the channel at different curvatures.
\end{abstract}

\section{Introduction}

Yield stress fluids are found in many situations, from toothpaste to drilling muds \cite{mousse13,barnes1989}. The property of the yield stress is often important, for example in preventing fluid flow in the absence of applied forces, and on the other hand often complicates applications, for example in requiring large stresses to be applied before a contaminated sludge can be processed. In varicose vein schlerotherapy \cite{nastasa2015} and enhanced oil recovery \cite{farajzadeh2012}, the yield stress of a foam allows it to act as a displacement fluid, pushing blood or oil respectively in front of it. In such applications it is necessary to predict in which parts of the fluid the stress will exceed the yield stress, and the material will flow, and where the stress is so low that either the material does not move at all, or moves as a solid plug.

Perhaps the simplest example of a continuum model for a yield stress fluid is due to Bingham \cite{bingham1922},  almost a century ago. This model assumes zero strain rate below a critical value of the stress and a linear relationship between stress and strain-rate above this value. This model has been extensively studied theoretically, for example for steady pressure-driven flow in straight channels of different cross-sections  \cite{birddy83,taylorw97,norouzivdbs15} and for boundary-driven flow in annuli~\cite{birddy83,muravlevemm10} and numerically, for example for flow past a sphere~\cite{blackerym97}.

Much of this work is concerned with Couette flow, as in a Couette viscometer, in which the fluid is held between concentric cylinders and one of the cylinder moves tangentially, Away from the laboratory, many flows of yield stress fluids are pressure-driven, often in curved or bent pipes~\cite{speddingbm04}. To the best of our knowledge, closed form analytic solutions for pressure-driven flow in an annulus have not been previously derived. 
Norouzi et al. \cite{norouzivdbs15} give an infinite series solution for the velocity profile in a curved three-dimensional channel with a rectangular cross-section. We take a different approach and, for simplicity, consider the equivalent 2D case, but seek a closed form expression for the velocity and stress profiles.

We consider the slow 2D pressure-driven (Poiseuille) flow of a Bingham fluid in a curved duct. The Dean number is assumed to be small (``creeping'' Dean flow), so that we neglect centripetal forces and any consequent secondary flows. In \S \ref{sec:maths} we give the governing equations of the flow, the constitutive equation for the fluid, and outline our solution, which requires determination of the radial  positions  of the yield surfaces, which we do numerically. We describe predictions for the velocity and stress fluids in \S \ref{sec:results}, and discuss the implications of our work for flow in narrow curved channels in \S \ref{sec:concs}.

\section{Mathematical model}
\label{sec:maths}

\begin{figure}[ht]
\centerline{
\includegraphics[width=0.5\textwidth]{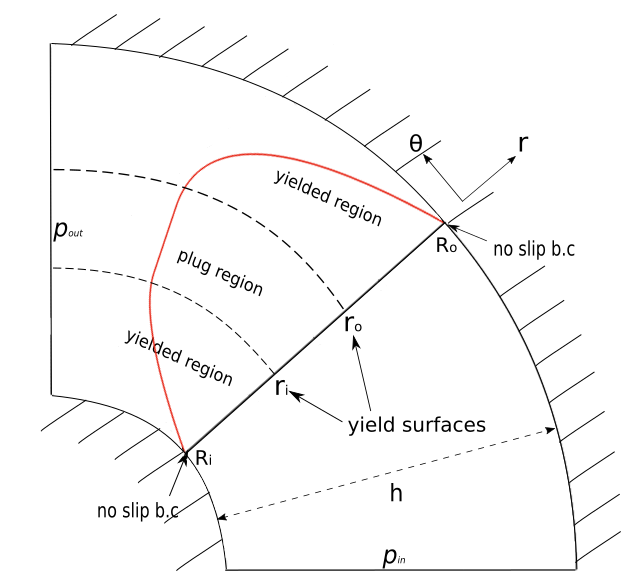}
}
\caption{The diagram indicates the geometry of the channel under consideration. Relative to plane polar coordinates ($r$, $\theta$), the channel has inner radius $R_i$ and outer radius $R_o$. Fluid flows in the positive $\theta$ direction due to a pressure difference $p_{\rm in} - p_{\rm out}$. An example of a velocity profile is shown in red, with a plug region between yield surfaces at $r=r_i$ and $r=r_o$.}
\label{fig:setup}
\end{figure}

\subsection{Governing equations}

We consider the steady flow of a Bingham fluid in the annular channel shown in Fig.~\ref{fig:setup}. The annulus has inner and outer radii  $R_i$ and $R_o$ respectively, giving a channel of width  $h=R_0-R_i$. 

The fluid moves in response to a constant pressure gradient $G$ acting in the $\theta$-direction, which can be written in terms of the inlet and outlet pressures $p_{in}$ and $p_{out}$ and the arc-length of the centreline of the channel, $R_{c} = \frac{1}{2} \left(R_i + R_o\right)$. 

Due to the existence of the yield-stress $\tau_0$ we must consider two distinct regions in the flow.
In the centre of the channel, where the shear stress $\tau_{r\theta}$ is small, there is a plug region of fluid moving at constant angular speed $u_\theta$. At the wall we impose a no-slip boundary condition, $u_\theta(R_i) = u_\theta(R_o) = 0$, so that at each side of the plug region, close to both walls of the channel, the strain-rate is greater and the fluid yields and flows like a Newtonian fluid with viscosity $\mu$.

We consider the constitutive equation in dimensionless form relative to the length-scale $R_c$ and a velocity scale $U_{m}$.
In this way the fluid's yield-stress $\tau_0$ and viscosity $\mu$ are absorbed into a dimensionless Bingham number $B$. The constitutive equation is then
\begin{equation} 
\label{eq:constitutive}
\begin{array}{ll}
 \tau_{r \theta}  = B + r \displaystyle \frac{\partial}{\partial r}  \left(\displaystyle \frac{u_{\theta}}{r} \right)  & \quad \text{for} \quad |\tau_{r \theta}| > B \\
\dot{\gamma}_{r \theta} = 0 & \quad \text{for} \quad |\tau_{r \theta}| \leq B,
\end{array}
\end{equation}
where $\dot{\gamma}_{r \theta}$ is the appropriate component of the strain-rate tensor. 

The  Bingham number is the ratio of the yield stress to the viscous stresses \cite{bingham1922}:
\begin{equation}
B = \frac{\tau_0 R_c}{\mu U_{m}}.
\end{equation}
For small values of $B$, the profiles of velocity and stress will be similar to those for a Newtonian fluid of comparable viscosity. Increasing $B$, for a fixed pressure gradient, allows a widening plug region to develop in the centre of the channel and results in a decrease in the fluid flux.

We consider slow, steady flows, and so the Stokes equations are the appropriate equations of motion~\cite{birddy83}. In the annular geometry these take the form
\begin{equation} \label{eq:stokes}
-\frac{R_{c}}{r} G =  \frac{1}{r^2} \frac{\partial}{\partial r} \left( r^2 \tau_{r \theta} \right).
\end{equation}
Note that the pressure gradient $G$ is scaled by the length-scale $R_c$ \cite{norouzivdbs15, rieger1973creeping} to take into account that the distance between the ends of the annular region increases with $r$, so that the pressure gradient should decrease with radial position.

\subsection{Analytic solution}

Equation~(\ref{eq:stokes}) has solution
\begin{equation}
\tau_{r\theta} = -\frac{G R_c}{2}  +  \frac{C}{r^2},
\label{eq:stress1}
\end{equation}
where $C$ is a constant of integration. Therefore the shear stress decreases quadratically across the gap, taking its maximum value at the inner wall $r= R_i$. This is because the pressure gradient is greatest on the inner wall.

In the absence of a yield stress, $B = 0$, the velocity profile for a Newtonian fluid in this geometry is
\begin{equation}
u_{\theta}(r) = \frac{GR_c}{2 \left(R_o^2 - R_i^2\right)}
 \left(
      R_i^2 r \ln\left(  \frac{r}{R_i}\right)  - R_o^2 r \ln \left(  \frac{r}{R_o}\right) 
  + \frac{R_i^2 R_o^2}{r} \ln \left(  \frac{R_i}{R_o}\right) \right) .
\label{eq:velNewt}
\end{equation}
This profile provides a reference for the more general case.

According to the constitutive equation for a Bingham fluid, eq.~(\ref{eq:constitutive}), the fluid yields when the magnitude of the shear stress is equal to the yield stress $B$. We can therefore find the positions of the inner and outer yield surfaces, $r_i$ and $r_o$, where $\tau_{r\theta} = B$ and $-B$ respectively:
\begin{equation}
r_i^2 = \frac{2C}{G R_c + 2B} , \quad  r_o^2 = \frac{2C}{G R_c -2 B}.
\end{equation}
Eliminating $C$ gives a relationship between the positions of the inner and outer yield surfaces, written in terms of modified Bingham numbers $B^{\pm} =  \frac{1}{2} G R_c \pm B $:
\begin{equation} \label{eq:yieldsurfrel}
B^{+} r_i^2 = B^{-} r_o^2 .
\end{equation}
In addition, substituting for $C$ in eq.~(\ref{eq:stress1}) gives two equivalent expressions for the stress in terms of the position of either yield surface:
\begin{equation} \label{eq:stress}
\tau_{r\theta} = -\frac{G R_c}{2} + B^+ \frac{{r_i}^2}{r^2}  \quad \mbox{and} \quad  
\tau_{r\theta} = -\frac{GR_c}{2} + B^- \frac{{r_o}^2}{r^2}
\end{equation}

This expression for the stress can now be substituted into the constitutive equation (\ref{eq:constitutive}) to give an expression for the velocity profile in each of the two yielded regions of the flow, $R_i \le r \le r_i$ and $r_o \le r \le R_0$. Between these regions the fluid moves in a solid-like plug, and so we assume a velocity profile  $u_\theta \propto r$ (solid body rotation) for $r_i \le r \le r_o$.
Applying no-slip boundary conditions at the walls $r=R_i$ and $r=R_o$, acknowledging the relationship between two yield surfaces i.e equation (\ref{eq:yieldsurfrel}) and prescribing that the velocity profile is continuous where the regions meet, at  $r=r_i$ and $r=r_o$, gives the velocity profile itself:
\begin{equation} 
\label{eq:velprof}
u_{\theta}(r) = \left\{
\begin{array}{lr}
B^+ \left( -r \ln\left(\displaystyle\frac{r}{R_i}\right) +\displaystyle \frac{r_i^2}{2} \left(\displaystyle \frac{r}{R_{i}^2} -\displaystyle\frac{1}{r} \right)  \right) & \quad \text{for} \quad R_i \leq r \le r_i \\
 B^+ r \left( -\ln\left(\displaystyle\frac{r_i}{R_i}\right) + \displaystyle\frac{1}{2} \left(\displaystyle \frac{r_i^2}{R_{i}^2} - 1 \right) \right) & \quad \text{for}\quad   r_i \leq r \leq r_o \\
B^- \left( - r \ln\left(\displaystyle\frac{r}{R_o}\right) - \displaystyle\frac{r_o^2}{2} \left(\displaystyle\frac{1}{r} - \displaystyle\frac{r}{R_{o}^2} \right)  \right) & \quad \text{for} \quad r_o \le r \leq R_o 
\end{array}
\right.
\end{equation}
as well as two conditions to determine the positions $r_i$ and $r_o$  of the yield surfaces:
\begin{equation} \label{eq:yieldsurf}
\begin{aligned}
B^{+} \left(\ln\left(\frac{r_i}{R_i}\right) + \frac{1}{2} \left( 1 - \frac{r_i^2}{R_i^2} \right) \right)  & =  B^{-} \left( \ln \left(\sqrt{\frac{B^{+}}{B^{-}}} \frac{{r_i}}{R_o} \right) + \frac{1}{2} \left( 1 - \frac{B^{+}}{B^{-}} \frac{{r_i}^2}{R_o^2}\right) \right), \\ B^{-} \left(\ln\left(\frac{r_o}{R_o}\right) + \frac{1}{2} \left( 1 - \frac{{r_o}^2}{R_o^2} \right) \right)  & =  B^{+} \left( \ln \left(\sqrt{\frac{B^{-}}{B^{+}}} \frac{{r_o}}{R_i} \right) + \frac{1}{2} \left( 1 - \frac{B^{-}}{B^{+}} \frac{{r_o}^2}{R_i^2}\right) \right).
\end{aligned}
\end{equation}

In order to find the yield surface positions we must solve this pair of nonlinear equations. This provides the necessary input to give the velocity profile $u_\theta(r) $ in equation~(\ref{eq:velprof}) in terms of the channel geometry ($R_i, R_o$) , the pressure gradient $G$ and the modified Bingham number $B^\pm$.

We solve the first condition in eqs.~(\ref{eq:yieldsurf}) for $r_i$ with Python using numpy's \textit{fsolve} root-finding algorithm, with initial guess $r_i = R_c$ (although the solution is the same for any initial guess in the range $[R_i,R_o]$).  Then the value of $r_0$ is determined from eq.~(\ref{eq:yieldsurfrel}).

\subsection{Constraints on the solution}

A Bingham fluid flowing through a straight channel of width $h$ due to a pressure gradient $G$ will flow provided that $G$ is sufficiently large. That is, below a critical pressure gradient the shear stress induced at the walls of the channel will not exceed the yield stress, and then the material will not move. The critical value in this case is found when the yield surfaces at $y = \pm B/G$ reach the walls at $y = \pm h/2$, and so can be expressed in the form $2B/Gh = 1$. (This quantity is referred to as the wall shear stress $T_0$ in \cite{birddy83}.)

In the curved channel  that we consider here, an indication that flow can cease is that for the velocity profile to be defined in eq.~(\ref{eq:yieldsurf}) we must have $B^-$ positive, giving an upper bound for the Bingham number, $ B \leq \frac{1}{2} G R_c$, or (equivalently) a lower bound for the pressure gradient.
 We write the ``yield ratio'' as $Y = 2B/Gh$, and seek the dependence of its critical value $Y_c$ on the channel curvature. For $Y< Y_c$ the yield surfaces always exist and the material flows.
 
Again using the fact that when the fluid is stationary the yield surfaces coincide with the walls of the channel, $r_i = R_i$ and $r_o = R_o$, we can derive the critical yield ratio $Y_c$ for the curved channel from equation~(\ref{eq:yieldsurfrel}).  Noting that $R_o = R_i + h$, we expand $B^{\pm}$ to find
\begin{equation}
\left( B + \frac{G}{4}(2R_i + h) \right) R_i^2 = \left( -B + \frac{G}{4}(2R_i + h)\right) (R_i + h)^2.
\end{equation}
This can be rearranged 
to give an expression for the critical yield ratio:
\begin{equation} \label{eq:crit2B/Gh}
Y_c =  1 - \frac{h^2}{4 R_i^2 + 4 h R_i + 2 h^2}.
\end{equation}
For large values of $R_i$, i.e. small channel curvature, $Y_c$ tends to one and we recover the result for the straight channel. 
For fixed $R_i$, increasing the channel width at high curvatures reduces the value of $Y_c$ significantly, although never below one half. 
Thus a small amount of channel curvature has a surprisingly large effect on inhibiting flow.

\section{Results}
\label{sec:results}

In presenting our results we use the yield ratio $Y$, scaled by its critical value $Y_c$, to represent the magnitude of the yield stress and a scaled radial position   $\hat{r} = (r - R_i)/h$, relative to the inner wall of the channel, to allow us to compare different values of the channel width $h$ and radius of curvature.

\subsection{Velocity}
\label{sec:results_velocity}

\begin{figure}
\centering
\subfloat[]{{\includegraphics[width=7.5cm]{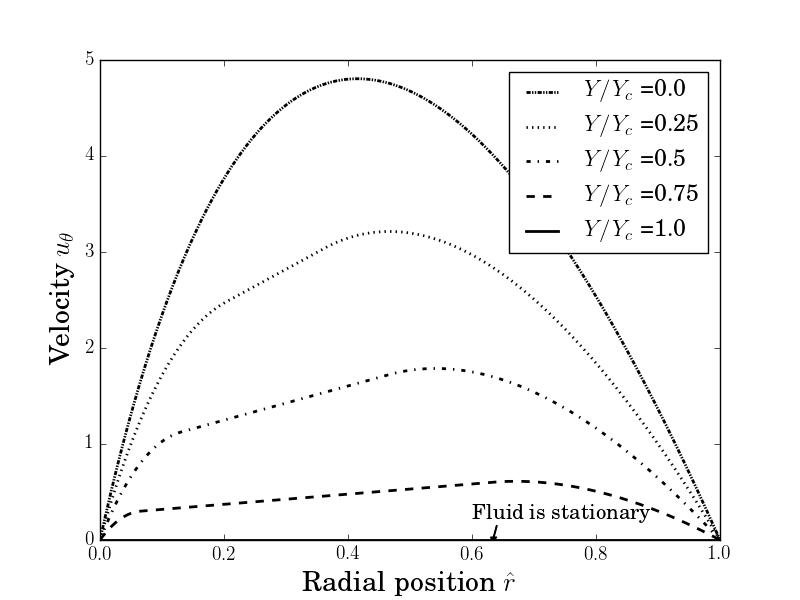} }}%
\qquad
\subfloat[]{{\includegraphics[width=7.5cm]{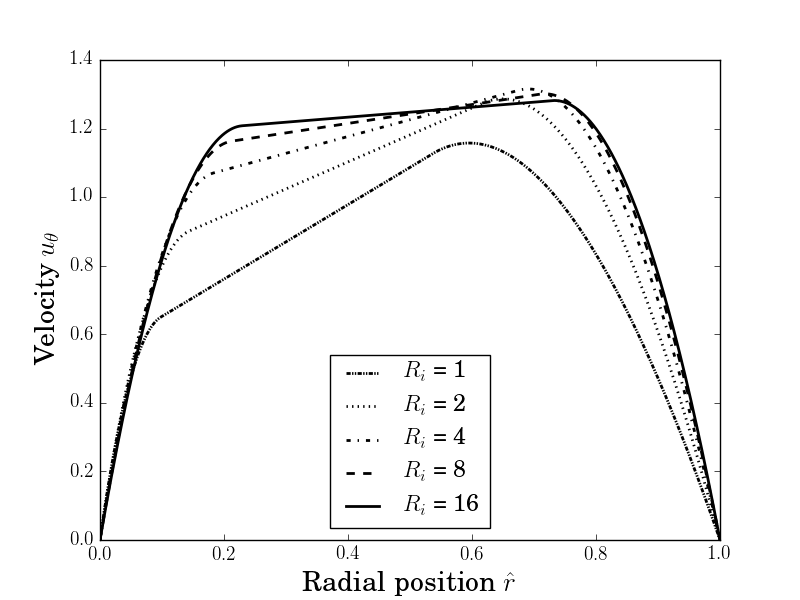} }}%
\caption{
Velocity profile in a curved channel as a function of radial position:
(a) for fixed channel geometry $R_i = 1$, $h=2$ and different  yield ratios.
(b) for constant yield ratio $Y/Y_c = 0.625$ and channel width $h=2$ and different values of the  inner radius.
}%
\label{fig:velocity}%
\end{figure}

For a given Bingham number $B$, pressure gradient $G$ and channel geometry, we can find the roots of eqs. (\ref{eq:yieldsurf}) for the  positions of both yield surfaces, described below (section \ref{sec:results_surface}) and hence plot the velocity  profile, eq. (\ref{eq:velprof}), shown in Fig. \ref{fig:velocity}.
Increasing the yield ratio $Y$ with a fixed channel geometry can be thought of as increasing the Bingham number $B$ or reducing the  pressure gradient $G$.  The result (Fig. \ref{fig:velocity}(a)) is that the fluid velocity decreases and the plug width increases. As $Y$ approaches $Y_c$, the flow stops.

Fig. \ref{fig:velocity}(b) shows how the velocity profile is affected by changes in channel curvature with fixed fluid properties $Y$.  Reducing the radius $R_i$ (and hence increasing the channel curvature) reduces the velocity, particularly in the inner half of the channel. 
The slope of the velocity in the plug region is high for small $R_i$ corresponding to high curvature of the stress profile.

The point of maximum velocity, close to $r=r_o$, moves away from the outer wall as $R_i$ decreases, but the value of the velocity there does not change monotonically: for intermediate curvature (e.g. $R_i = 4$) the maximum velocity of the fluid  exceeds the value  in a straight channel.

\begin{figure}
\centering
\subfloat[]{{\includegraphics[width=7.5cm]{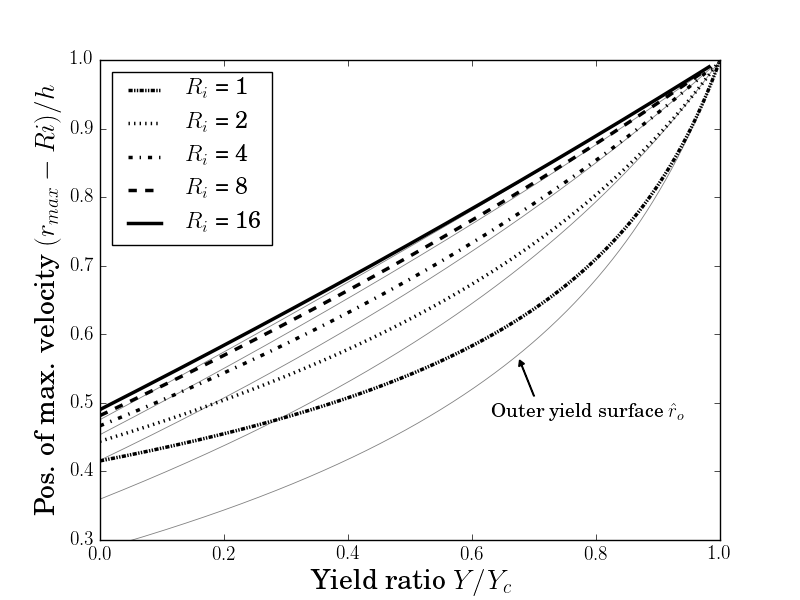} }}%
\qquad
\subfloat[]{{\includegraphics[width=7.5cm]{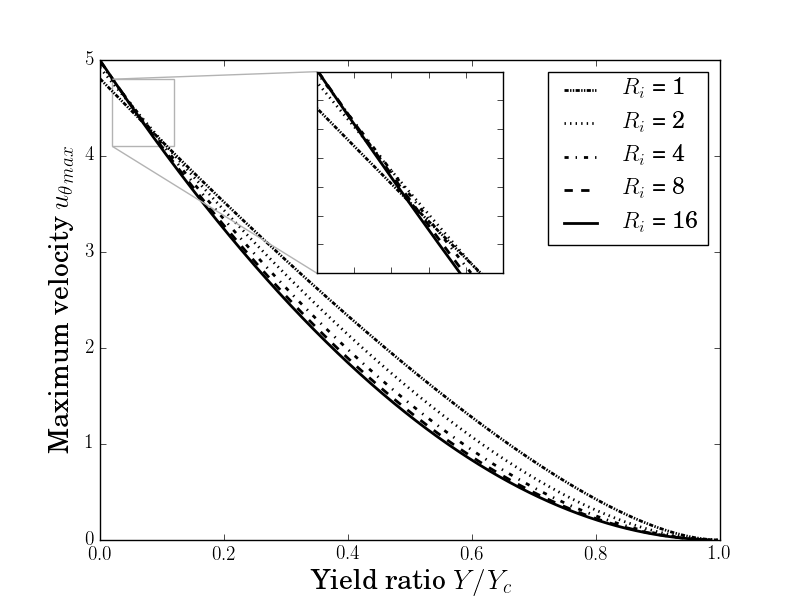} }}%
\caption{
The point of maximum velocity with fixed channel width $h=2$ as a function of $Y/Y_c$. 
(a) The radial position of the maximum velocity (thick lines), compared with the position of the outer yield surface (thin lines) for several values of the inner radius $R_i$.
(b) The value of the maximum velocity when $r = r_{max}$.
 }%
\label{fig:maxvelpos}%
\end{figure}

Figure~\ref{fig:maxvelpos}(a) shows that the radial position of the point of maximum velocity is always greater than or equal to $r_o$. Differentiation of the velocity profile (eq.~\ref{eq:velprof}) appropriate to radial positions $r_o \leq r \leq R_o$ gives:
\begin{equation}
\frac{d}{dr}(u_\theta(r)) = -B^{-} \left( \ln\left(\frac{r}{R_o}\right) + 1 - \frac{r_o^2}{2}\left(\frac{1}{r^2} + \frac{1}{R_o^2}\right)\right).
\end{equation}
Equating this to zero gives an expression for the radial position of maximum velocity $r_{max}$:
\begin{equation}
\ln\left(\frac{r_{max}}{R_o}\right) + 1 = \frac{r_o^2}{2} \left( \frac{1}{r_{max}^2} + \frac{1}{ R_o^2} \right).
\end{equation}
A numerical solution is shown in Fig. \ref{fig:maxvelpos}(a) for different $R_i$ as the yield ratio $Y$ changes. As $R_i$ increases (decreasing channel curvature), the profile of position approaches a linear interpolation between the midpoint of the channel for $Y=0$ and the outside of the channel for $Y=Y_c$, as for a straight channel.
As $Y$ increases relative to $Y_c$, the slope of the velocity profile in the plug region is reduced (and the fluid velocity also decreases). Hence $r_{max}$ approaches the outer yield surface and they eventually coalesce (Fig. \ref{fig:maxvelpos}(a)). For less curved channels (large $R_i$, this coalescence is seen at smaller values of $Y/Y_c$.

The value of the maximum velocity itself, ${u_{\theta}}_{max}$, is shown in Fig. \ref{fig:maxvelpos}(b) for different $R_i$ as the yield ratio $Y$ changes.
For $Y = 0$, i.e. a Newtonian fluid, the maximum velocity of the fluid increases as the channel curvature decreases (larger $R_i$). There is a value of $Y/Y_c \approx 0.084$ at which the maximum velocity appears independent of channel curvature, although it occurs at a different position for each value of $R_i$.Then for larger values of $Y/Y_c$, a channel with greater curvature induces a flow with a  higher maximum velocity, and this point occurs at a radial position further away from the outer wall.

At the crossover, the yield-stress is small relative to the pressure gradient, but nonetheless indicates the competition between the curvature of the channel and the yield stress of the fluid in determining the motion. The position of the point of maximum velocity is far enough from the outer wall that the no slip condition is not dominant, but not so close to the inner wall that the higher curvature there induces larger stresses.

\subsection{Yield surface positions and plug width}
\label{sec:results_surface}

\begin{figure}
\centering
\subfloat[]{{\includegraphics[width=7.5cm]{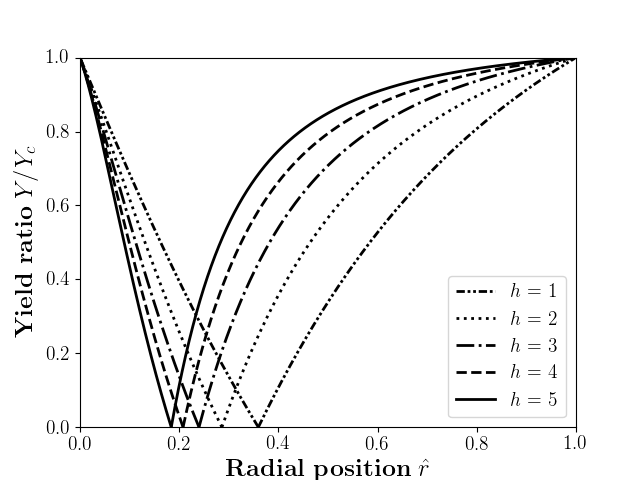} }}%
\qquad
\subfloat[]{{\includegraphics[width=7.5cm]{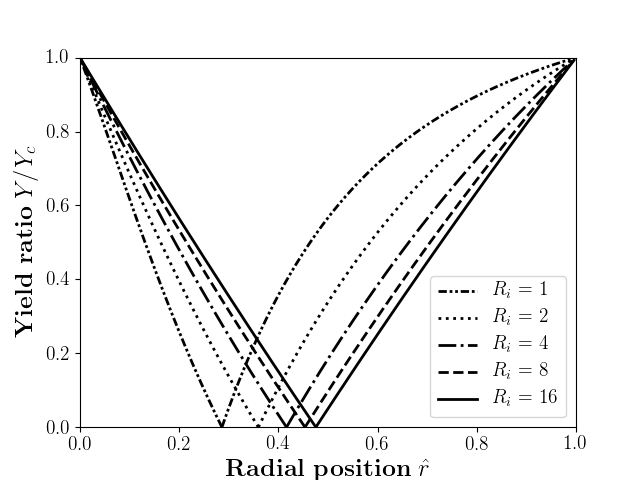} }}%
\caption{Yield surface positions as a function of yield ratio: (a) for different channel widths, with fixed inner wall radius $R_i = 1$; (b) for fixed channel width $h=2$ and different values of the inner radius.
}%
\label{fig:ys}%
\end{figure}

The solution of eqs.~(\ref{eq:yieldsurf}) for the radial positions of the yield surfaces is shown in Fig.~\ref{fig:ys}. 
In the limit  $Y \rightarrow 0$, the material behaves like a Newtonian fluid:  there are no yield surfaces and the values $r_i$ and $r_o$ coincide at a point close to the middle of the channel. 
As the width $h$ of the channel increases (Fig. \ref{fig:ys}(a)) or the radius of curvature of the channel decreases (Fig. \ref{fig:ys}(b)) this point moves towards the inner wall. As $R_i \rightarrow \infty$ they meet at $\hat{r} = \frac{1}{2}$, as in a straight channel.

As $Y$ increases, the two yield surfaces move apart, reaching the inner and outer walls precisely when $Y$ reaches $Y_c$.  For large channel widths $h$ and small inner radius $R_i$ the outer yield surface remains close to the centre of the channel until $Y$ reaches about half of its critical value, while the position of the inner yield surface is almost linear in $Y/Y_c$ in all cases.

\begin{figure}
\centering
\subfloat[]{{\includegraphics[width=7.5cm]{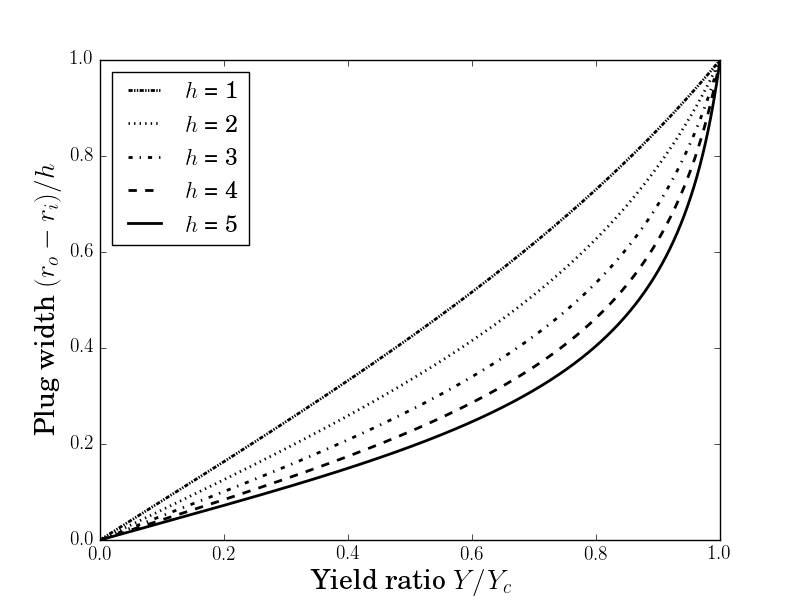} }}%
\qquad
\subfloat[]{{\includegraphics[width=7.5cm]{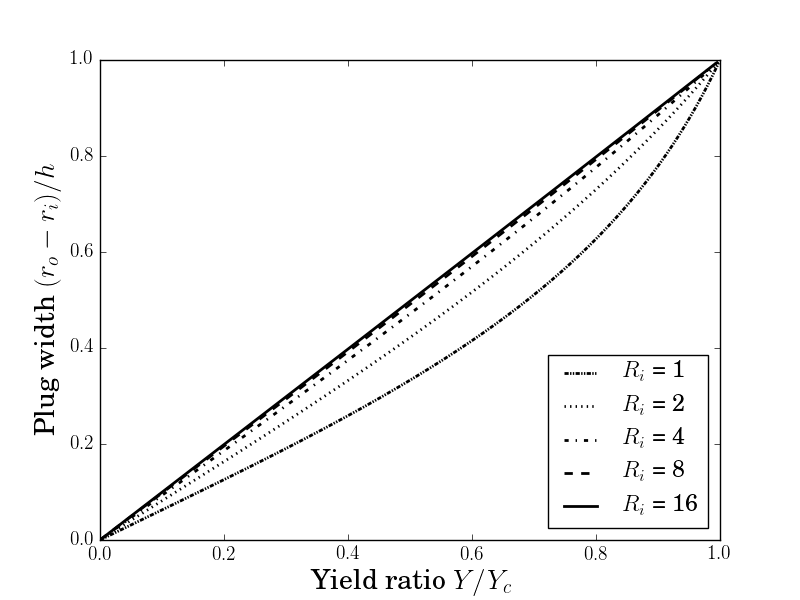} }}%
\caption{Plug width as a function of yield ratio: (a) for different channel widths, with fixed inner wall radius $R_i = 1$; (b) for fixed channel width $h=2$ and different values of the inner radius.
}%
\label{fig:pluglength}%
\end{figure}

The distance between the yield surfaces is the plug width, the region of low stress in which the material moves as a solid body, shown in Fig.~\ref{fig:pluglength}.  As  $Y  \rightarrow Y_c $ the plug width increases until it spans the whole channel.  As the width of the channel increases, the plug width increases more slowly with $Y$ (Fig.~\ref{fig:pluglength}(a)). The effect of varying the radius of the inner cylinder $R_i$ (Fig.~\ref{fig:pluglength}) is weaker, but at high curvature (small $R_i$) there is a slight decrease in the plug width for given $Y$. For weakly curved channels (large $R_i$), the plug width becomes linear in $Y/Y_c$ (Fig.~\ref{fig:pluglength}(b)).

\subsection{Shear stress}

\begin{figure}
\centering
\subfloat[]{{\includegraphics[width=7.5cm]{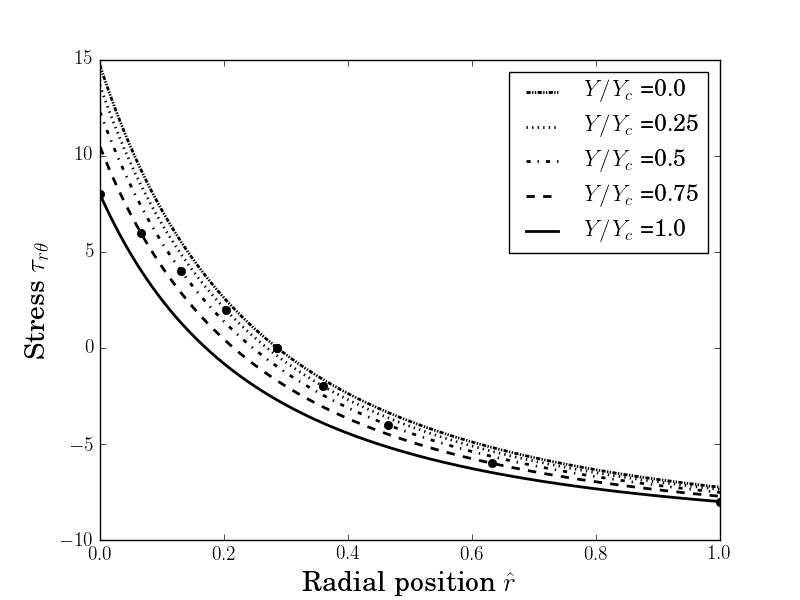} }}%
\qquad
\subfloat[]{{\includegraphics[width=7.5cm]{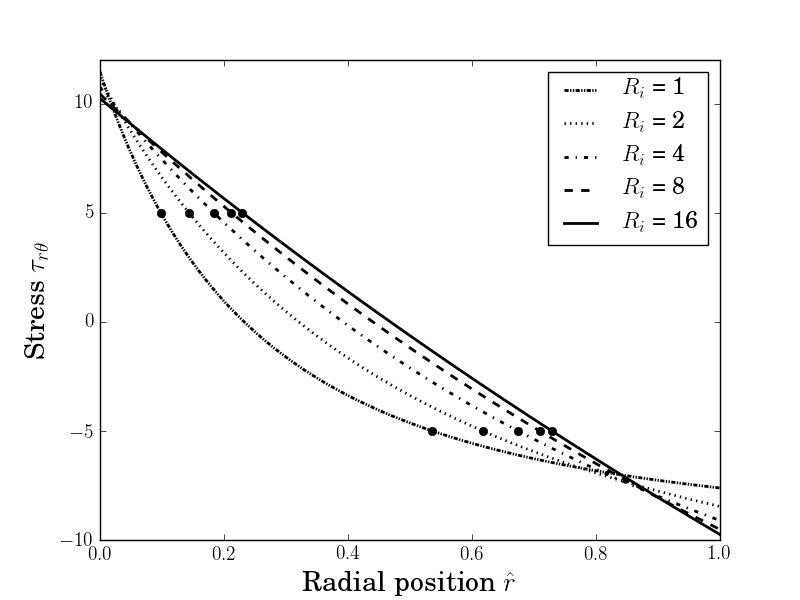} }}%
\caption{
Stress profiles as a function of radial position, with the position of the yield surfaces, at which the magnitude of the stress  is equal to the Bingham number $B$, marked with dots. 
(a) For a fixed channel geometry with $h = 2$ and $R_i = 1$,  with different yield ratios.
 (b) For fixed $Y/Y_c = 0.625$ and channel width $h = 2$, corresponding to $B=5$, with different values of the inner radius.
 }%
\label{fig:stress}%
\end{figure}

Equation~(\ref{eq:stress1}) shows that the shear stress decreases from the inner to the outer wall, since the pressure gradient is greatest at the inner wall. 
Figure \ref{fig:stress}(a) shows the profile of stress in a channel with $R_i = 1, h=2$ for different values of the yield ratio $Y$.
As $Y$ increases towards $Y_c$, the stress decreases everywhere. The maximum value of stress, located at the inner wall, reduces towards $\tau_{r\theta} = B$, while the stress at the outer wall decreases only slightly.

The effect on the stress of varying the inner radius $R_i$, is significant (Fig.~\ref{fig:stress}(b)). As $R_i$ increases, the plug width increases and the stress profile becomes straighter, approaching the linear profile found in a straight channel.

\subsection{Flux}

A useful quantity to predict is the amount of fluid that flows through the channel. We calculate the one-dimensional flux $Q$, i.e. the amount of fluid which crosses a particular cross-section per unit time, by integrating the velocity profile (eq.~(\ref{eq:velprof})) with respect to radial position:
\begin{equation}
\begin{split}
Q &= \int_{R_i}^{R_o} u_{\theta} {\rm d}r  \\
  &= \int_{R_i}^{r_i} u_{\theta} {\rm d}r  +  \int_{r_i}^{r_o} u_{\theta} {\rm d}r +  \int_{r_o}^{R_0} u_{\theta} {\rm d}r  \\
&= B^{+} \left( \frac{{r_i}^2}{2}\left(\ln\left(\frac{R_i}{r_i} \right) + \frac{1}{2}\right) + \frac{{r_o}^2}{2}\left(\ln\left(\frac{R_i}{r_i}\right) - \frac{1}{2}\right) -\frac{R_i^2}{4} + \frac{r_o^2 r_i^2}{4R_i^2}\right) +B^{-} \left( r_o^2 \ln\left(\frac{{r_o}}{R_o}\right) + \frac{R_o^2}{4} - \frac{{r_o}^4}{4 R_o^2} \right)
\end{split}
\end{equation}

Recognising that the flux is the area beneath the velocity curves in Fig.~\ref{fig:velocity}, we  expect $Q$ to tend to zero as the yield ratio approaches its critical value $Y_c$, while for $Y = 0$ the flux $Q$ is the value for a Newtonian fluid responding to the same pressure gradient.

\begin{figure}
\centering
\subfloat[]{{\includegraphics[width=7.5cm]{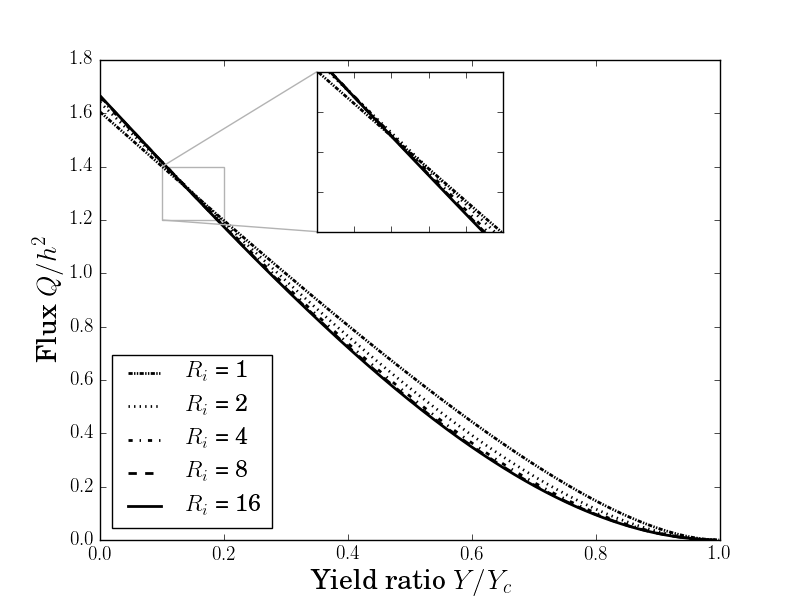} }}%
\qquad
\subfloat[]{{\includegraphics[width=7.5cm]{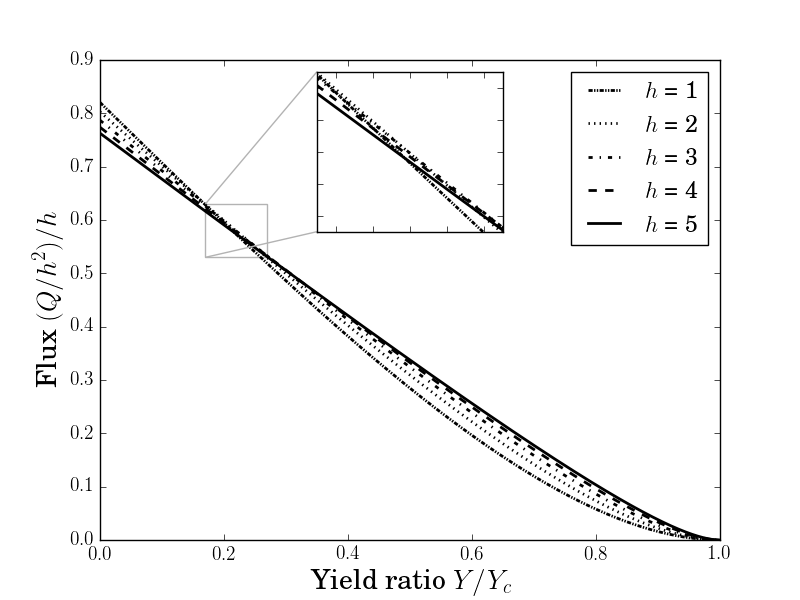} }}%
\caption{The flux through the channel as a function of yield ratio $Y$: 
(a) for fixed channel width $h =2$ and different values of the inner radius.
(b) for different channel widths, with fixed inner radius $R_i = 1$. 
}%
\label{fig:flux}%
\end{figure}

We report the flux as an area of fluid passing through the channel per unit time, so scale it by $h^2$ (Fig.~\ref{fig:flux}(a)). At low Bingham number, the flux is greatest for large $R_i$: increasing curvature of the channel reduces the amount of material moving through the channel for given width $h$.
However, just as for the maximum velocity (Fig.~\ref{fig:maxvelpos}), at a moderate value of $Y/Y_c$ close to 0.15 there is a crossover, and the flow through a curved channel is greater for given $Y/Y_c$. (Note that $Y_c$ depends on the channel geometry (eq.~(\ref{eq:crit2B/Gh}) so this is not equivalent to an increase of flux due to increases channel curvature for fixed $B$ and $G$.)

For variable channel width $h$ we further scale the flux with an additional factor of channel width (Fig.~\ref{fig:flux}(b)).
There is again a crossover, at $Y/Y_c \approx 0.21$, where the flux $(Q/h^2)/h$ is similar for all channel widths. 
For any yield ratios which are less than than this critical value, the wider channels produce a relatively smaller flux. This is reversed once for larger values of $Y/Y_c$.

\section{Discussion \& Conclusion}
\label{sec:concs}

Poiseuille flows of yield stress fluids in curved channels have attracted relatively little attention. It is clear that the scenario we describe, of a pressure-driven flow in a curved channel, is difficult to implement in an experiment in isolation. Instead, it could be thought of as one element of, for example, a network of pipes conveying some yield stress fluid, which due to certain constraints must be made to turn a corner.

Our predictions allow  the effect of such a situation to be determined, for example the drop in flux associated with such a bend, as a function of the material parameters of the fluid. This work also provides a more stringent test against which to validate simulation codes for rheological models in non-trivial geometries and as a base flow which is perturbed when the flow-rate increases and secondary flows may arise.

Our main result, eq. (\ref{eq:velprof}), provides detailed insight into the dependence of the flow on the dimensions of the channel. It allows us to identify non-monotonicity in the flow, in particular in the region of maximum velocity (Fig.~\ref{fig:maxvelpos}(b)), stress (Fig.~\ref{fig:stress}(b)) and flux (Fig.~\ref{fig:flux}) in the channel.

In terms of the scaled yield ratio $Y/Y_c$, which depends on the ratio of the Bingham number $B$ and the pressure gradient $G$, we consider two extreme situations. For small $Y$, the fluid behaves like a Newtonian fluid, with relatively large velocity and large stress on the inner wall $R_i$. Such a material is likely to be ineffective at displacing a second fluid (in the example of varicose vein treatments, this second fluid is the blood that initially fills the vein), because it will be prone to instabilities such as viscous fingering.

In the other limit, as $Y/Y_c \rightarrow 1$, the flow is dominated by the yield stress of the fluid and is relatively slow. The majority of the material move as a large plug which almost completely spans the channel (Fig.~\ref{fig:pluglength}). In applications, it is this plug region which is essential for displacing another fluid. So a large Bingham number, relative to the applied pressure gradient, is required in varicose vein schlerotherapy, for example, to empty the vein and allow it to collapse. 

Our result also indicates that the degree of curvature of the channel affects the efficacy of a displacement flow. For each yield ratio $Y/Y_c$, the width of the plug region is smaller for channels with greater curvature. In the varicose vein example, a vein that is manipulated in such a way to reduce its curvature should be treated more effectively.

\section*{Acknowledgements}

We acknowledge financial support from the UK Engineering and Physical Sciences Research Council (EP/N002326/1) and a PhD studentship from BTG.

\bibliography{bingham_curved_05.bib} 
\bibliographystyle{unsrt}

\end{document}